\documentclass[12pt]{article}

\usepackage[letterpaper,hmargin=1in,vmargin=1in]{geometry}

\usepackage{graphicx,epstopdf,amsmath,amsfonts,amssymb,cite}
\usepackage[hypertex]{hyperref}

\def\be{\begin{equation}}
\def\ee{\end{equation}}
\def\ba{\begin{eqnarray}}
\def\ea{\end{eqnarray}}
\def\ge{\mathrel{\raise.3ex\hbox{$>$\kern-.75em\lower1ex\hbox{$\sim$}}}}
\def\la{\mathrel{\raise.3ex\hbox{$<$\kern-.75em\lower1ex\hbox{$\sim$}}}}

\def\simgt{\mathrel{\raise.3ex\hbox{$>$\kern-.75em\lower1ex\hbox{$\sim$}}}}
\def\simlt{\mathrel{\raise.3ex\hbox{$<$\kern-.75em\lower1ex\hbox{$\sim$}}}}

\newcommand{\bi}[1]{\bibitem{#1}}
\newcommand{\fr}[2]{\frac{#1}{#2}}

\newcommand{\nc}{\newcommand}

\nc{\gone}{\bar g_{\pi NN}^{(1)}}
\nc{\gzero}{\bar g_{\pi NN}^{(0)}}
\nc{\al}{\alpha}
\nc{\ga}{\gamma}
\nc{\de}{\delta}
\nc{\ep}{\epsilon}
\nc{\ze}{\zeta}
\nc{\et}{\eta}
\nc{\ka}{\kappa}
\nc{\rh}{\rho}
\nc{\si}{\sigma}
\nc{\ta}{\tau}
\nc{\up}{\upsilon}
\nc{\ph}{\phi}
\nc{\ch}{\chi}
\nc{\ps}{\psi}
\nc{\om}{\omega}
\nc{\Ga}{\Gamma}
\nc{\De}{\Delta}
\nc{\La}{\Lambda}
\nc{\Si}{\Sigma}
\nc{\Up}{\Upsilon}
\nc{\Ph}{\Phi}
\nc{\Ps}{\Psi}
\nc{\Om}{\Omega}
\nc{\ptl}{\partial}
\nc{\del}{\nabla}
\nc{\ov}{\overline}
\nc{\newcaption}[1]{\centerline{\parbox{15cm}{\caption{#1}}}}
\nc{\us}{U(1)$_S$}

\def\beq{\begin{equation}}
\def\eeq{\end{equation}}
\def\bmat{\begin{displaymath}}
\def\emat{\end{displaymath}}
\def\bear{\begin{eqnarray}}
\def\eear{\end{eqnarray}}
\def\ba{\begin{eqnarray}}
\def\ea{\end{eqnarray}}
\def\bery{\begin{array}}
\def\ery{\end{array}}
\def\bit{\begin{itemize}}
\def\eit{\end{itemize}}
\def\ben{\begin{enumerate}}
\def\een{\end{enumerate}}
\def\btab{\begin{tabular}}
\def\etab{\end{tabular}}
\def\btbl{\begin{table}}
\def\etbl{\end{table}}
\def\bfig{\begin{figure}[htb]}
\def\efig{\end{figure}}
\def\bpic{\begin{picture}}
\def\epic{\end{picture}}

\def\ga{\mathrel{\raise.3ex\hbox{$>$\kern-.75em\lower1ex\hbox{$\sim$}}}}
\def\la{\mathrel{\raise.3ex\hbox{$<$\kern-.75em\lower1ex\hbox{$\sim$}}}}
\def\gappeq{\mathrel{\rlap {\raise.5ex\hbox{$>$}}
{\lower.5ex\hbox{$\sim$}}}}
\def\lappeq{\mathrel{\rlap{\raise.5ex\hbox{$<$}}
{\lower.5ex\hbox{$\sim$}}}}

\def\gyr{{\rm \, G\kern-0.125em yr}}
\def\mev{{\rm \, Me\kern-0.125em V}}
\def\gev{{\rm \, Ge\kern-0.125em V}}
\def\tev{{\rm \, Te\kern-0.125em V}}

\begin{document}

\begin{titlepage}

\setcounter{page}{1}

\vspace*{0.2in}

\begin{center}

\hspace*{-0.6cm}\parbox{17.5cm}{\Large \bf \begin{center}

\boldmath{$B_s$} Mixing and Electric Dipole Moments in MFV

\end{center}}

\vspace*{0.5cm}
\normalsize

\vspace*{0.5cm}
\normalsize

{\bf Brian Batell$^{\,(a)}$ and  Maxim Pospelov$^{\,(a,b)}$}

\smallskip
\medskip

$^{\,(a)}${\it Perimeter Institute for Theoretical Physics, Waterloo,
ON, N2J 2W9, Canada}

$^{\,(b)}${\it Department of Physics and Astronomy, University of Victoria, \\
     Victoria, BC, V8P 1A1 Canada}

\smallskip
\end{center}
\vskip0.2in

\centerline{\large\bf Abstract}

We analyze the general structure of four-fermion operators capable of introducing 
$CP$-violation preferentially in $B_s$ mixing within the framework of Minimal Flavor Violation. 
The effect requires a minimum of $O(Y_u^4Y_d^4)$ Yukawa insertions, and at this order we find a total of six 
operators with different Lorentz, color, and flavor contractions that lead to enhanced $B_s$ mixing. 
We then estimate the impact of these operators 
and of their close relatives on the possible sizes of electric dipole moments (EDMs) of neutrons and heavy atoms. 
We identify two broad classes of such operators: those that 
give EDMs in the limit of vanishing CKM angles, and those that require quark mixing for the 
existence of non-zero EDMs. The natural value for EDMs from the operators in the first category 
is up to an order of magnitude above the experimental upper bounds, while the second group predicts 
EDMs well below the current sensitivity level. Finally, we 
discuss plausible UV-completions for each type of operator.

\vfil
\leftline{June 2009}
    
\end{titlepage}

\subsection*{1. Introduction}

Studies of the $B_d$ mesons in the last decade \cite{Review} have
confirmed the economical Cabibbo-Kobayashi-Maskawa (CKM) mechanism of flavor and $CP$-violation 
implicit in the Standard Model (SM), described entirely by three mixing angles and a single phase.
Despite a continuing improvement in precision, a practical question arises as to 
the best strategy to look for new flavor/$CP$-violating effects
in $B$-systems. One formidable opportunity is presented by the $B_s$ system, where the 
$CP$-violation due to the SM CKM phase is naturally small, so that a large amount of $CP$ violation 
would necessarily be attributable to a New Physics (NP) source. In other words, $CP$ violation in 
the $B_s$ system belongs to the same CKM-background-free category of tests as electric dipole moments 
(EDMs) of neutrons and heavy atoms.

It is therefore intriguing that recent tests of $CP$ violation in the decays of $B$ mesons at
the Tevatron experiments show a deviation from the predictions of the SM. Specifically, both 
the CDF and D0 experiments have reported correlated measurements of the width difference 
$\Delta \Gamma_s$ and $CP$-violating phase $\phi_s^{J/\psi\phi}$ from an analysis of the angular 
distributions of the flavor-tagged decays $B^0_s \rightarrow J/\psi \phi$. 
Early results~\cite{jp1,jp2} showed a deviation with the SM at the $2.1\sigma$ level~\cite{jp3}, 
while a more recent preliminary measurement~\cite{jp4} is consistent with the SM, albeit still with 
relatively large error bars. More recently, the D0 collaboration announced \cite{D0ass} the 
measurement of the same-sign dimuon 
asymmetry in $B$-decays:
\begin{equation}
A^b_{\rm sl} \equiv \frac{N_b^{++} - N_b^{--}}{N_b^{++} + N_b^{--}}
  = -(9.57 \pm 2.51 \pm 1.46) \times 10^{-3},
\end{equation}
where $N_b^{++}$ is the number of events in which two antimuons are produced from the decays of a $B$ 
and $\bar{B}$ meson, {\rm i.e.} $b\bar{b} \rightarrow \mu^+ \mu^+ X$. This is to be compared to the SM 
prediction, $A^b_{\rm sl}({\rm SM}) = (−2.3^{+0.5}_{-0.6}) \times 10^{−4}$ \cite{Grossman,LN}, and thus shows a 
$3.2\sigma$ deviation from the predicted value. 
Given the tighter constraints in the well-studied $B_d$ system, 
this measurement therefore also hints at a NP source of $CP$ violation in  $B_s$ mixing. Indeed, interpreted at face value, 
the D0 result \cite{D0ass} implies the presence of a $CP$-violating part of the $B_s-\bar B_s$,
\be
\label{ImM}
{\rm Im} (\Delta M_{12}) \sim O(|M_{12}^{\rm SM}|) \sim O(10\,{\rm psec}^{-1}).
\ee
We refer the reader to the detailed numerical fits of the data recently performed in Refs.~\cite{Buranow,Jurenow}.
We focus on NP contributions to $\Delta M_{12}$, but see \cite{Gamma} for a recent analysis of the possible NP contributions
to $\Gamma_{12}$.
Perhaps the ultimate test of $CP$ in $B_s$ systems will be delivered by the LHCb experiment in the near future. 

If these recent measurements from the Tevatron experiments are to be ascribed to a NP source of $CP$ violation, 
a natural question is how such NP contributes to EDMs. On the experimental front, significant progress has been 
achieved in the measurement of $^{199}$Hg EDM, where the upper bound has been improved by a factor of 7 \cite{Hg},
\be
\label{dHg}
|d_{Hg}| < 3.1\times 10^{-29}~e{\rm cm}.
\ee
This considerably tightens all bounds on flavor-conserving hadronic and semi-leptonic $CP$-violating operators, 
and in particular implies rather strong bounds on the color EDMs of quarks. 

In this note we investigate a possible connection between $CP$ violation in 
$B_s$ mixing and EDMs within the framework of 
Minimal Flavor Violation (MFV) \cite{MFV1,MFV2}. This framework assumes that NP, 
should it induce flavor change, preserves the re-parametrization independence of the 
SM flavor physics. In other words, the flavor transitions are governed by 
the CKM matrix $V_{CKM}$ and the eigenvalues of the Yukawa couplings, but new 
$CP$-violating phases introduced in a flavor-universal way are allowed \cite{susyMFV1,susyMFV2}.  
We identify all $\Delta F=2$ four-fermion operators leading to the preferential 
introduction of $CP$ violation in the $B_s$ system.
To leading order in Yukawa insertions, the required operators arise at $O(Y_u^4Y_d^4)$. 
These are specific realizations of the class of operators pointed out in
Ref.~\cite{generalMFV}, where the general consequences of Yukawa 
insertions at any order were investigated. The origin of $CP$ violation is flavor-blind, 
and the enhancement of its effect in $B_s$ relative to $B_d$ is governed by the ratio of 
Yukawa couplings $y_s/y_d$ \cite{generalMFV}.
Notice that despite the fact that the total number of Yukawa insertions is rather large, the 
effect is not necessarily hopelessly small simply because the scale of the Yukawa coupling 
in the down-type sector is very uncertain. In particular, the Yukawa couplings of $b$-quark may not 
lead to any suppression at all, $y_b \sim O(1)$, as happens in large $\tan\beta$ two-Higgs doublet models. 
An example of a model in this category would be the large $\tan\beta$ supersymmetric model,
where even in the limit of very heavy superpartners 
the Higgs exchange leads to important effects both in flavor physics \cite{SUSYgrave} and 
EDMs \cite{LP,DLOPR}. For an MFV vs EDM discussion, see \cite{EDMMFV}. See also Ref.~\cite{RS} for 
an early study of SUSY effects on lepton asymmetries in $B$ systems. 

Normalizing the size of these $CP$-violating four-fermion operators to a putative 
signal in $B_s$ decays, {\em i.e.} to the maximal size consistent with 
the mass splitting, we next address the question of EDMs. We point out that 
all these operators and their close relatives can be further subdivided into two broad classes. 
The first class is contains the scalar-pseudoscalar Lorentz structure $S\times P$ and survives 
in the limit of $V_{CKM}\to 1$. The neutron EDM is predicted typically close to experimental 
bounds and the natural size of the mercury EDM is up to one order of magnitude above the experimental 
limits. The second class of operators has the structure of left- times right-handed vector current, 
$(V-A)\times(V+A)$. To have $CP$ violation, this class of operators requires more than one generation, 
and as a result all EDMs acquire additional suppression by $V_{ts}^2$ or $V_{td}^2$, bringing them 
well below the sensitivity of modern EDM experiments. We provide examples of possible UV 
completions for both types of operators. 

In Section 2 we present a detailed classification of these operators, along with the estimate of their
plausible size that makes them detectable in the $B_s$ mixing. Section 3 provides some 
background information on EDMs and estimates the contribution from each operator to $d_{Hg}$ and $d_n$. 
Finally, Section 4 contains a discussion of possible UV completions and presents our conclusions. 

\subsection*{2. $CP$-violating MFV operators for the $B_s$ mixing}

The rules of the game in MFV are defined by the requirement to retain the existing flavor
re-parametrization freedom of the SM. Since the right-handed rotations remain unphysical, we use 
this freedom to put all Yukawa couplings to the Hermitian form
\be
Y_u^\dagger = Y_u,~~ Y_d^\dagger = Y_d,
\ee
which allows us for shorter expressions. 

A four-fermion operator relevant for $B_s$ mixing can be written in the following general form,
\be
\label{bsbs}
O = (\bar b \Gamma_{A} s) (\bar b \Gamma'_{A} s), 
\ee
where $\Gamma^{(')}_{A}$ stands for all possible Lorentz and color contractions. 
Lifting such operators to the flavor space according to the 
MFV rules, one encounters two types of MFV operators:
those that give identical $CP$-violating NP contributions in both $B_d$ and $B_s$ mixing in comparison to their 
SM values, and those that are enhanced in $B_s$ mixing \cite{generalMFV}. We are interested in the 
latter, and therefore specify the following criteria for selecting the operators:
\begin{enumerate}

\item $O$ violates $CP$ and contains $\Delta F =2$ transitions. 

\item The contribution of $O$ to the mixing in $B_s$ is enhanced over $B_d$ by $y_s/y_d$. 

\item $O$ survives the limit of $y_c^2,~y_u^2 \to 0$, and involves no more than 
one power of $y_s$ or $y_d$. 

\item  $O$ is a local Lorentz scalar mediated by the exchange of particles heavier than $m_B$. 

\item The total number of Yukawa insertions in $O$ is minimized.

\end{enumerate} 

Needless to say, we also require an overall electric and color neutrality of $O$, but
do not impose any $SU(2)\times U(1)$ requirements as those can be easily satisfied 
by an appropriate number of Higgs insertions. 
While the first four conditions on the list are crucial for our discussion, the last one is 
for book-keeping purposes only, as any $Y_u^2$ insertions can 
be taken as $(Y_u^2)^n$, resulting in a proliferation of powers of top quark Yukawa,
but not bringing any additional numerical smallness. The condition of locality explicitly 
forbids a situation where the $B$-mixing is mediated by a neutral particle with a
mass close to $m_{B_s}$ or $m_{B_d}$, which can enhance the mixing in either of these two systems 
via a resonance, with a possibility of distorting MFV relations. 

When generalizing (\ref{bsbs}) to the full flavor space, one has to remember that 
flavor indices can be contracted outside of a quark pair that has its Lorentz/color indices contracted.  
We choose to eliminate such operators using completeness in the flavor sector, but then have to 
account for all possible Lorentz and color structures. 
Calling the left- and right-handed down-quarks as $Q$ and $D$, 
with the above guiding conditions at hand, we arrive at the following 
set $\{ O_s\}$ of the effective $CP$-violating operators:

\begin{eqnarray}
\label{Os}
O_1 & = & i (\bar Q^k \; Y_u^2Y_d  \;D^k)  \;\; (\bar D^l \; Y_d [Y_d^2,Y_u^2]  \;Q^l) 
+(h.c.),\\\nonumber
O_2 & = &  i (\bar Q^k \; Y_u^2Y_d  \;D^l)  \;\; (\bar D^l \; Y_d [Y_d^2,Y_u^2]  \;Q^k) 
+(h.c.),\\\nonumber
O_3   & =  & i (\bar D^k \;Y_d [ Y_d^2, Y_u^2] Y_d \gamma_\mu \;D^k)  \;\; (\bar Q^l \;Y_u^2 \gamma_\mu \;Q^l), \\\nonumber
O_{4} & = & i (\bar D^k\; Y_d [ Y_d^2, Y_u^2] Y_d \gamma_\mu \;D^l) \;\;  (\bar Q^l \;Y_u^2 \gamma_\mu \;Q^k),  \\\nonumber
O_{5} & = & i (\bar D^k \;Y_d  Y_u^2 Y_d \gamma_\mu \;D^k) \;\;  (\bar Q^k \;[Y_d^2,Y_u^2] \gamma_\mu  \;Q^l),\\\nonumber
O_{6} & = & i (\bar D^k \;Y_d  Y_u^2 Y_d \gamma_\mu \;D^l) \;\;  (\bar Q^l \;[Y_d^2,Y_u^2] \gamma_\mu  \;Q^k).
\end{eqnarray}

In these expressions, $[,]$ 
denote commutators in the 
flavor space, superscripts $k$ and $l$ show the contraction of color $SU(3)$ indices while 
the Lorentz and flavor indices are contracted within each parentheses. From the point of view of 
$\Delta F = 2 $ flavor transitions, the set $\{ O_s\}$ is clearly over-complete. Indeed, {\em e.g.} 
to leading order in the strange quark Yukawa coupling
several of them lead to the same $(\bar b_L s_R)  (\bar b_R  s_L)$ or $(\bar b_L \gamma_\mu s_L)
(\bar b_R\gamma_\mu s_R)$ operators. 
However, we should not count these operators as a priori redundant as they might contain 
differences in $\Delta F =0 $ channels and thus have different manifestations in the EDMs. 
All of these operators contain commutators in the flavor space and 
therefore vanish in the limit of $V_{CKM} \to 1$. 
While the 
flavor structures in $O_{3}-O_{6}$ are clearly dictated by 
the properties of Hermiticity, the choice of flavor structure in $O_1$ and $O_2$ 
is not unique. One can take, for example, $i(\bar Q^k \; Y_u^2Y_d  \;D^k)  \; (\bar D^l \; Y_d^3Y_u^2 \;Q^l)$, 
which also leads to $CP$ -violation in $B_s$ mixing. However, this can be reduced to 
$O_1$ since a structure analogous to $O_1$ with an anti-commutator instead of commutator 
gives a vanishing contribution to $\Delta F = 2$ operators. 

Choosing the normalization constant to be $G_F/\sqrt{2}$, we combine these 
operators into an effective $CP$-odd Lagrangian weighted with dimensionless 
coefficients $c_i$:
\be
\label{Leff}
{\cal L}^{CP} = \fr{G_F}{\sqrt{2}}\sum_{i=1..6} c_i O_i.
\ee
Next, we reduce Eq.~(\ref{Leff}) to the subset of operators leading to 
$\Delta F = 2$ transitions for $B$-mesons, 
finding two independent structures for each light flavor:
\be
{\cal L}_{\Delta F = 2}^{CP} = i\fr{G_F}{\sqrt{2}} y_t^4y_b^3 {V_{tb}^*}^2 
\sum_{q=d,s}y_q  V_{tq}^2\left [ C_{SLR} (\bar b_L q_R) (\bar b_R q_L)  + C_{VLR} (\bar b_L \gamma_\mu q_L)
(\bar b_R\gamma_\mu q_R) \right] +(h.c.),
\ee
where the Wilson coefficients are related to the original classification as follows:
\be
 C_{SLR} = 2(c_1 - c_4 - c_6) ;~~ C_{VLR} = c_3 + c_5 - c_2
\ee

These operators should be evolved using perturbative QCD from the scale where they are generated 
to the $B$-meson energy scale. It is hard to do this in general since we do not know the actual scale 
where these operators are generated. A reasonable assumption is that this scale is rather large, 
comparable to the EW scale, in which case we can directly use the results of QCD evolution and the 
calculated matrix elements  already present in the literature
(see {\em e.g.} \cite{BurasQCD}). This produces the following estimate of the $CP$-odd mixing part of $B_s$. 
\begin{eqnarray}
\label{M12}
{\rm Im} ( M_{12} ) & \simeq  &  \fr{G_F}{\sqrt{2}}\fr{ m_{B_s} F_{B_s}^2}{3} y_t^4y_b^3y_s 
|V_{tb}^*  V_{ts} |^2\times (P_1 C_{VLR} + P_2 C_{SLR})\\
& \simeq & ( 10 {\rm psec}^{-1})\times \fr{y_b^3y_s}{10^{-3}}
\left[ c_1 - c_4 - c_6  +0.33\left(c_2-c_3 - c_5 \right)    \right],
\nonumber
\end{eqnarray}
where $P_1 \simeq -1.62$ and $P_2 \simeq 2.46$ are from Ref. \cite{BurasQCD}. The eigenvalues of Yukawa 
matrices are normalized at a high-energy scale.  We have disregarded the small complex phase of 
$V_{tb}^*  V_{ts}$, and took this product to be equal to 0.04, and $y_t \simeq 1$. 
The overall coefficient in (\ref{M12}) is chosen to be very close to  half of the measured absolute value of 
$\Delta M_{B_s}$.

Besides the presence of six unknown Wilson coefficients, an additional uncertainty in the
estimate (\ref{M12}) is the value of 
the combination of Yukawa couplings from the down-quark sector. In the SM such a combination is 
hopelessly small, but at large $\tan\beta$ this combination can be as large as $10^{-2}$, 
so that (\ref{M12}) will cause large 
effect in $B_s$ mixing. It should also be said that at very large $\tan\beta$ the relation 
between measured masses and eigenvalues of the Yukawa couplings, {\em e.g.} $y_s/y_b \simeq m_s/m_b$ 
weakens considerably because of the possibility of very large corrections to the 
mass operator \cite{AnotherSUSYgrave}. The result (\ref{M12}) shows that there is some room 
for the generation of $\{ O_s \}$ at the weak scale with nearly maximal $\tan\beta$, 
a point emphasized recently in Ref.~\cite{DFM}. 

Current experiments are capable of detecting only the maximal amount of $CP$-violating $B_s$ mixing \cite{reach1,reach2,reach3}. 
Therefore, as a benchmark,  we choose to equate the the imaginary part of the NP contribution to 
$\Delta m_{B_s}/2 \simeq 10$ psec$^{-1}$. This benchmark fixes the combination of $y_b^3y_s$ times 
the linear combination of Wilson coefficients. We shall now use this as an approximate input for 
the estimates of the natural size of EDMs in this framework. 

\subsection*{3. Natural size of EDMs from $\{ O_s \}$ and its extensions}

Electric dipole moments of heavy atoms and neutrons (see Refs. \cite{FG,PR} for reviews) is a 
powerful probe of new $CP$-violating physics at and above the weak scale. EDMs do not 
require flavor transition and therefore may be induced by NP even in the limit of
$V_{CKM} \to 1$. It is remarkable that despite the fact that MFV has extra flavor-universal
$CP$-violating phases,  {\em all} 
operators in the set $\{O_s\}$ vanish at $V_{CKM} \to 1$. 
Moreover, it turns out that operators $O_1$ and $O_2$ 
do not contain $CP$-violating terms for $\Delta F=0$ processes. 
Indeed, the flavor commutator requires the presence of 
quarks from two different generations, {\em e.g.} $s$ and $b$, 
which makes such operators $\propto (\bar b_L s_R) (\bar s_R b_L)$.
These in turn can be Fierz-transformed to the products of $s$- and $d$- 
vector and axial vector currents, $(\bar s_R  \gamma_\mu s_R) (\bar  b_L \gamma_\mu b_L)$, that
always conserve  $CP$. Retaining only $y_b^3y_{s(d)}$-proportional contributions, 
we choose to eliminate all $(V-A)\times(V+A)$ operators 
with Fierz transformations, 
arriving at the following $\Delta F=0$ component of the effective Lagrangian (\ref{Leff}):
\be
\label{F=0}
{\cal L}^{CP}_{\Delta F =0} = i\fr{G_F}{\sqrt{2}}  y_t^4y_b^3 
\sum_{q=d,s}2y_q|V_{tb}V_{tq}|^2\left [ (c_4-c_6)(\bar b_L^k b_R^k)(\bar q_R^l q_L^l)   
+ (c_3-c_5) (\bar b_L^k b_R^l)(\bar q_R^l q_L^k) \right] + (h.c.). 
\ee
Thus, even before estimating the actual size of the EDMs, we can conclude that
there exist natural choices of operators within MFV, such as $O_1$ and $O_2$, that contribute 
to $CP$-violation in $B_s$ but do not lead to EDMs. 

Turning to the actual size of the EDMs induced by (\ref{F=0}), we expect them to be very small
on account of the additional suppression by $|V_{td}|^2 \sim 10^{-4}$. 
There are several pathways for the Lagrangian (\ref{F=0}) to contribute to EDM observables. 
Integrating out the $b$-quark induces EDMs $d_q$ and 
color EDMs $\tilde d_q$ of light quarks that in turn 
lead to a neutron EDM, as well as $CP$-odd
nuclear forces that manifest in 
$d_{Hg}$. It is easy to see that ${\cal L}^{CP}_{\Delta F =0}$
mixes with $d_q$ and $\tilde d_q$ only at two-loop level.

A good proxy for the strength of
$CP$-odd nuclear  forces is given by the $CP$-odd pion-nucleon and $\eta$-nucleon coupling  constants. 
Besides being induced by the two-loop $\tilde d_q$, these couplings also 
receive a more direct contribution from (\ref{F=0}) \cite{LP,Barr} 
via the ``heavy quark content'' of a nucleon,
$\langle N | m_b \bar bb | N\rangle \simeq 65$ MeV. 
Taking the $\bar b b \bar d i \gamma_5 d$ part of ${\cal L}^{CP}_{\Delta F =0}$, 
we estimate the strength of the pion-nucleon coupling constant to be
\begin{eqnarray}
\label{gpiNN} 
\label{tree}
g_{\pi NN} = (c_4 -c_6) \fr{G_F}{\sqrt{2}} |V_{tb}V_{tq}|^2 y_t^4y_b^3 y_d \times \fr{\eta_{QCD} R_{QCD}}{f_\pi},
\end{eqnarray}
where $R_{QCD}$ is a factorized combination of QCD condensates and nucleon matrix elements 
of quark bi-linears:
\be
R_{QCD} = \langle N | \bar bb | N \rangle \langle 0 |  \bar dd| 0 \rangle -
\langle N | \bar dd | N \rangle \langle  0| \bar bb | 0 \rangle \sim 5\times 10^{-4} ~({\rm GeV})^3.
\ee
The numerical value of $R_{QCD}$ is obtained assuming 
the following values of the quark condensates and matrix elements: 
$\langle N | \bar bb | N \rangle \sim 1.35\times 10^{-2}$, 
$\langle 0| \bar bb | 0 \rangle = - \langle G_{\mu\nu}^{a2} \rangle \alpha_s
/(12\pi m_b) \sim - ( 55 ~{\rm MeV} )^3 $, 
$\langle N | \bar dd | N \rangle \sim 4$, and 
$\langle 0 |  \bar dd| 0 \rangle \sim -(250 {\rm MeV})^3$. 
There is also an enhancement coefficient $\eta_{QCD}\sim 2.5$ due to QCD running of the 
$\bar b_L b_R \bar d_R d_L$ operator from the UV scale down to the 
scale $m_b$. 
All of these steps produce the following estimate of the pion-nucleon coupling constant:
\be
g_{\pi NN} \sim 5 \times 10^{-16} \times \fr{y_b^3y_s}{10^{-3}} (c_4 -c_6),
\ee
where we also used $y_d/y_s\sim 0.04$.

The two-loop results for the color EDMs can be obtained by ``recycling'' the 
calculations of the Barr-Zee diagrams for the 2HDM in the limit 
of $m_b^2 \ll m_{\rm Higgs}^2$ \cite{BZ,Chang,HPR}: 
\begin{eqnarray} 
\tilde d_d^{~2-loop} & = & \fr{\alpha_s m_b}{16\pi^3} \fr{G_F}{\sqrt{2}} 
|V_{tb}V_{tq}|^2 y_t^4y_b^3 y_d \times \ln\frac{\Lambda_{UV}}{m_b} \times \left(c_4 -c_6+\fr76(c_3-c_5)\right)\\
 & \simeq &    \fr{y_b^3y_s}{10^{-3}} \left(c_4 -c_6+\fr76(c_3-c_5)\right) \times 5\times 10^{-30} ~{\rm cm}.
\label{2loop}
\end{eqnarray} 
In the second relation we also used $\Lambda _{UV} \sim G_F^{-1/2}$ and chose $\alpha_s\sim 0.15$.
This result should be compared with the 
limit on the $CP$-odd pion-nucleon coupling in
the isospin $=1$ channel extracted from the mercury EDM \cite{Hg}
and the implied limit on color EDMs of quarks \cite{Hg,FG,Pospelov}: 
\be
\label{limit}
|g^{1}_{\pi NN}| < 10^{-12};~~  |\tilde d_d - \tilde d_u| < 6 \times 10^{-27}~{\rm cm} \, .
\ee 
It shows that the choice $y_b^3y_sc_i \sim O(10^{-3})$ motivated by maximal 
$CP$-violation in $B_s$ mixing (\ref{M12}) yields EDMs that are 
three orders of magnitude below the experimental bounds. Even taking into account significant theoretical 
uncertainties involved in estimates (\ref{tree}) and (\ref{2loop}) as well as nuclear/QCD 
uncertainties in extracting (\ref{limit}) from $d_{Hg}$, one can 
conclude that the EDMs induced by operators from $\{O_s\}$  are well 
below current levels 
as well as anticipated future experimental
sensitivity benchmarks.

One of the main reasons why the results (\ref{tree}) and (\ref{2loop}) 
are so small is the strong suppression coming from the factor $|V_{td}|^2 \simeq 10^{-4}$,
which is a consequence of the commutators in flavor space present in every member
of $\{ O_s\}$. In this respect, it is reasonable to investigate whether close 
``flavor relatives'' of $O_i$ give EDMs in the limit $V_{CKM}\to 1$. 
We define a flavor relative as a modification of an operator $O_i$ where some flavor-structure is 
added/removed within each quark pair in a way consistent with MFV. 
For example, a ``minimal'' flavor relative of $O_1$ would be the operator
$ (\bar Q^k Y_d D^k)  (\bar D^l Y_d Q^l) $.
It is easy to see that all operators $O_3-O_6$ that involve a 
product of left- and right-handed currents do not have flavor 
relatives that give large EDMs. Indeed, removing 
any of the  $Y_u^2$ or $Y_d^2$ insertions
leads to operators that conserve $CP$. Therefore, flavor relatives 
of $O_3-O_6$ always require $V_{CKM}\neq 1$ to induce EDMs, and these EDMs are small
according to (\ref{tree}) and (\ref{2loop}). On the contrary, 
there exists flavor relatives of $O_1$ and $O_2$ that do give EDMs 
in the limit of $V_{CKM} \to 1$:
\be
O_{1,2}~\to~ i(\bar Q^{k}\;Y_d^3  \;D^{k})  (\bar D^l \; Y_d  \;Q^l) +(h.c.) \to 
iy_b^3y_d (\bar b_Lb_R)(\bar d_R d_L)+(h.c.).
\ee
If UV physics generates these relatives of $O_1$ and $O_2$ 
with similar size Wilson coefficients, the resulting EDMs 
are four orders of magnitude above (\ref{tree}) and (\ref{2loop}), on the order of 
$\tilde d_d \sim 5 \times 10^{-26}$ cm. This corresponds to EDMs 
right at the current level of experimental sensitivity for the neutron \cite{nedm} and about one order 
of magnitude above the current bounds for mercury. The effective field 
theory approach does not allow one to make a more refined statement before the UV physics is specified.

As a final comment in this section, it may still be possible that D0 same-sign dimuon asymmetry \cite{D0ass} 
has a non-negligible contribution from a NP source that does not differentiate between 
$B_s$ and $B_d$ by an extra factor of $y_s/y_d$, 
and for whatever reason the presence of NP in $B_d$ has not been detected elsewhere. 
In this case, the spectrum of 
$CP$-violating operators broadens rather considerably. Some of these operators, 
such as pure left-handed type $i(\bar Q [Y_u^2,Y_d^2] \gamma_\mu Q) (\bar Q Y_u^2 \gamma_\mu  Q)$
and its flavor relatives, cannot induce large EDMs. Others involve 
the chirality flip, $(\bar Q Y_u^2 Y_d D) (\bar Q Y_u^2 Y_d D)$, and their flavor/Lorentz relatives may induce 
significant EDMs even at one-loop level:
\be
i(\bar Q  Y_d \sigma_{\mu\nu} D) (\bar Q Y_d \sigma_{\mu\nu}D) \to
i y_by_d  (\bar b_L \sigma_{\mu\nu} b_R) (\bar d_L  \sigma_{\mu\nu} d_R) 
\to   d_d^{1-loop}.
\ee
If the combination $y_b y_d$ corresponds to a choice of maximal $\tan\beta$,
such one-loop EDMs will be very large, being enhanced 
relative to (\ref{2loop}) by $O(100 \times V_{td}^{-2})\sim 10^6$, 
and indeed several orders of magnitude above all EDM bounds. A detailed analysis of such operators 
falls outside the scope of the present paper.

\subsection*{4. Discussion }

In this paper we have presented the explicit form of the $CP$-odd $\Delta B = 2$ MFV operators
that predominantly contribute to the $CP$-violation of $B_s$ mesons \cite{generalMFV}. 
Since the $CP$-violating properties are governed by the flavor-universal
phase in front of these operators, one might have expected
large effects for EDMs. On the contrary, we have found that 
$\{ O_s \}$ requires non-zero CKM mixing matrix elements 
for the EDMs to exist. This extra $|V_{td}|^2$ suppression 
places EDMs directly induced by $\{O_s\}$ well below 
the experimental bounds. At the same time ``close flavor relatives'' of scalar operators 
$O_1$ and $O_2$ give EDMs on the order of $10^{-26}$ cm, comparable or even somewhat 
larger than the current best limits \cite{Hg}. On the other hand, operators of the type $O_3-O_6$ 
that involve the product of left- and right-handed currents $(V-A)\times (V+A)$
{\em do not have} flavor relatives that generate EDMs in the limit $V_{CKM}\to 1$. 
Our general analysis has implications for the specific UV completion schemes 
that may be responsible for generating operators $O_i$ that lead to preferential 
$CP$-violation in the $B_s$ system.

{\em 2HDM, MSSM, and color-octet scalars.} 
The exchange of MFV scalars \cite{Buranow,MW} can generate 
$O_1$ and/or $O_2$ operators and their flavor relatives, and there
is no good argument why EDMs should be small. In the minimal supersymmetric model (MSSM),
Higgs exchange at large  $\tan\beta$ combined with SUSY radiative corrections 
to the mass sector of down-type quarks can be a significant source of $\Delta F = 2$ operators. 
If $CP$-violation is introduced in an MFV-like fashion, it has to be sourced by the 
relative phase of the $\mu$-parameter in the superpotential and the gluino mass parameter. 
To have an effect on $B_s$, this phase would have to be nearly maximal. 
Besides the effects induced by $O_1$ and $O_2$ discussed in this paper, there 
will be of course the one-loop $\tan\beta$-enhanced EDMs that will be sensitive to the 
scale of the light scalar quark and lepton masses in excess of $\sim 5$ TeV. 
With the necessity to keep heavy Higgs masses under a TeV, the model would require 
a number of fine-tunings of different mass/phase parameters to survive the EDM constraints. 
If, however, all these tuning requirements are met and mass scales in the squark/slepton sector 
are pushed into multi-TeV range, while $m_{A(H)}$ are kept under a TeV, there are still 
non-vanishing contributions to atomic EDMs on account of hadronic and semileptonic 
four-fermion operators, {\em e.g.} $\bar qq \bar e i \gamma_5 e$ \cite{LP}. Given the analysis 
of this paper and of Ref. \cite{LP}, such EDMs will be at the current limits of 
experimental observability (or even slightly above current bounds).
It would be interesting to investigate $CP$-violation in the $B_s$ system and EDMs in SUSY models 
extended by neutral chiral superfields (NMSSM) on account of possibly light mediators and new 
sources for the $CP$-violating phases. 

{\em  Vector and pseudoscalar exchange.} The exchange by neutral vector particles such as $Z$-bosons (or hypothetical $Z'$)
is the most relaxed possibility with respect to the EDM  constraints because it results only in 
$O_3-O_6$ operators. The $Z$-boson, of course, does not have any flavor-changing couplings, and those would 
have to be generated by integrating out NP. 
An explicit example of such couplings consistent with MFV was given 
recently in models with extra vector-like quarks \cite{mike1}. These models may also 
have additional, and not necessarily 
small effects in $B_d$ and $K$-mesons from pure 
left-handed operators, $\bar Q Y_u^2 Q \bar Q Y_u^2 Q $,
and the associated CKM phase. 
The derivatively-coupled pseudoscalar particles, with couplings $f_a^{-1}\partial_\mu a \bar D \gamma_\mu D$ 
and alike, are in principle capable of inducing similar effects if $f_a$ is under a TeV, but on account of the extra 
derivative would necessarily have to be relatively light, $m_a \sim m_B$. 

{\em Exchange by particles transforming nontrivially under flavor.}
Lastly, the exchange by particles that transform non-trivially under 
flavor-rotations (see e.g. Refs.~\cite{mike2,mike1}) is also capable of 
inducing $\{ O_s\}$. In this case, however, the correspondence between spin of the mediators and our operator classification 
may be different than in the case of the mediators that transform trivially under flavor. 
For examples, exchange by scalars that transform as $(3,\bar 3)$ under the  $SU(3)_Q$ and $SU(3)_D$ 
would generate operators $\bar Q^f D^j \bar D^j Q^f$ where $j,f$ are flavor indices. After a Fierz transformation,
this operator obtains the $(V-A)\times (V+A)$ structure and therefore requires additional CKM suppression to induce EDMs.

\subsubsection*{Acknowledgements}

M.P. would like to thank the participants of the JHU workshop for a number of useful discussions. 
The work of M.P. is supported in part by NSERC, Canada, and research at the Perimeter Institute
is supported in part by the Government of Canada through NSERC and by the Province of Ontario through MEDT.


\begin{thebibliography}{99}


\bi{Review}  For a recent review, see {\it e.g.} A.~J.~Buras,
  Prog.\ Theor.\ Phys.\  {\bf 122}, 145 (2009)
  [arXiv:0904.4917 [hep-ph]].
 
\bibitem{jp1}
  V.~M.~Abazov {\it et al.}  [D0 Collaboration],
  Phys.\ Rev.\ Lett.\  {\bf 101}, 241801 (2008)
  [arXiv:0802.2255];

 \bibitem{jp2}
  T.~Aaltonen {\it et al.}  [CDF Collaboration],
  Phys.\ Rev.\ Lett.\  {\bf 100}, 161802 (2008)
  [arXiv:0712.2397 [hep-ex]].

\bibitem{jp3}
CDF/D\O\ $\Delta \Gamma_s$, $\beta_s$ Combination Working Group, D\O\ Note 5928-CONF (2009).

\bibitem{jp4}
L.~Oakes (CDF Collaboration), talk at FPCP 2010, May 25-29, Torino, Italy,
\url{http://agenda.infn.it/getFile.py/access?contribId=12&resId=0&materialId=slides&confId=2635}
 
\bibitem{D0ass}
V.~M.~Abazov {\it et al.}  [The D0 Collaboration],
arXiv:1005.2757 [hep-ex].
  
\bibitem{Grossman}
  Y.~Grossman, Y.~Nir and G.~Raz,
  Phys.\ Rev.\ Lett.\  {\bf 97}, 151801 (2006)
  [arXiv:hep-ph/0605028].

\bibitem{LN}
  A.~Lenz and U.~Nierste,
  JHEP {\bf 0706}, 072 (2007)
  [arXiv:hep-ph/0612167].
 
  \bibitem{Buranow}  A.~J.~Buras, M.~V.~Carlucci, S.~Gori and G.~Isidori,
  arXiv:1005.5310 [hep-ph].
 
  \bibitem{Jurenow}Z.~Ligeti, M.~Papucci, G.~Perez and J.~Zupan,
  arXiv:1006.0432 [hep-ph].
  
\bibitem{Gamma}
  A.~Dighe, A.~Kundu and S.~Nandi,
  arXiv:1005.4051 [hep-ph];
  C.~W.~Bauer and N.~D.~Dunn,
  arXiv:1006.1629 [hep-ph].


  \bibitem{Hg}
  W.~C.~Griffith, M.~D.~Swallows, T.~H.~Loftus, M.~V.~Romalis, B.~R.~Heckel and E.~N.~Fortson,
  Phys.\ Rev.\ Lett.\  {\bf 102}, 101601 (2009).

\bibitem{MFV1} 
  R.~S.~Chivukula and H.~Georgi,
  Phys.\ Lett.\  B {\bf 188}, 99 (1987);
  L.~J.~Hall and L.~Randall,
  Phys.\ Rev.\ Lett.\  {\bf 65}, 2939 (1990);
  A.~J.~Buras, P.~Gambino, M.~Gorbahn, S.~Jager and L.~Silvestrini,
  Phys.\ Lett.\  B {\bf 500}, 161 (2001)
  [arXiv:hep-ph/0007085].

\bibitem{MFV2} 
  G.~D'Ambrosio, G.~F.~Giudice, G.~Isidori and A.~Strumia,
  Nucl.\ Phys.\  B {\bf 645}, 155 (2002)
  [arXiv:hep-ph/0207036].

\bibitem{susyMFV1}
  J.~R.~Ellis, J.~S.~Lee and A.~Pilaftsis,
  Phys.\ Rev.\  D {\bf 76}, 115011 (2007)
  [arXiv:0708.2079 [hep-ph]].

\bibitem{susyMFV2}
  G.~Colangelo, E.~Nikolidakis and C.~Smith,
  Eur.\ Phys.\ J.\  C {\bf 59}, 75 (2009)
  [arXiv:0807.0801 [hep-ph]].

\bibitem{generalMFV}  A.~L.~Kagan, G.~Perez, T.~Volansky and J.~Zupan,
  Phys.\ Rev.\  D {\bf 80}, 076002 (2009)
  [arXiv:0903.1794 [hep-ph]].
  
  \bibitem{SUSYgrave} R.~Hempfling,
  Phys.\ Rev.\  D {\bf 49}, 6168 (1994); C.~Hamzaoui, M.~Pospelov and M.~Toharia,
  Phys.\ Rev.\  D {\bf 59}, 095005 (1999)
  [arXiv:hep-ph/9807350]; K.~S.~Babu and C.~F.~Kolda,
  Phys.\ Rev.\ Lett.\  {\bf 84}, 228 (2000)
  [arXiv:hep-ph/9909476].

\bi{LP}  O.~Lebedev and M.~Pospelov,
  Phys.\ Rev.\ Lett.\  {\bf 89}, 101801 (2002)
  [arXiv:hep-ph/0204359].

\bibitem{DLOPR}
  D.~A.~Demir, O.~Lebedev, K.~A.~Olive, M.~Pospelov and A.~Ritz,
  Nucl.\ Phys.\  B {\bf 680}, 339 (2004)
  [arXiv:hep-ph/0311314].
  
  \bi{EDMMFV} L.~Mercolli and C.~Smith,
  Nucl.\ Phys.\  B {\bf 817}, 1 (2009)
  [arXiv:0902.1949 [hep-ph]];  P.~Paradisi and D.~M.~Straub,
  Phys.\ Lett.\  B {\bf 684}, 147 (2010)
  [arXiv:0906.4551 [hep-ph]].

\bibitem{RS}
  L.~Randall and S.~f.~Su,
  Nucl.\ Phys.\  B {\bf 540}, 37 (1999)
  [arXiv:hep-ph/9807377].

  
  \bi{BurasQCD} A.~J.~Buras, S.~Jager and J.~Urban,
  Nucl.\ Phys.\  B {\bf 605}, 600 (2001)
  [arXiv:hep-ph/0102316].

\bi{AnotherSUSYgrave} 
L.~J.~Hall, R.~Rattazzi and U.~Sarid,
  Phys.\ Rev.\  D {\bf 50}, 7048 (1994)
  [arXiv:hep-ph/9306309];  T.~Blazek, S.~Raby and S.~Pokorski,
  Phys.\ Rev.\  D {\bf 52}, 4151 (1995)
  [arXiv:hep-ph/9504364]; C.~Hamzaoui and M.~Pospelov,
  Eur.\ Phys.\ J.\  C {\bf 8}, 151 (1999)
  [arXiv:hep-ph/9803354];  B.~A.~Dobrescu and P.~J.~Fox,
  arXiv:1001.3147 [hep-ph].


\bi{DFM} 
  B.~A.~Dobrescu, P.~J.~Fox and A.~Martin,
  arXiv:1005.4238 [hep-ph].

\bibitem{reach1}
  S.~Laplace, Z.~Ligeti, Y.~Nir and G.~Perez,
  Phys.\ Rev.\ D {\bf 65}, 094040 (2002)
  [hep-ph/0202010].

\bibitem{reach2}
  M.~Beneke, G.~Buchalla, A.~Lenz and U.~Nierste,
  Phys.\ Lett.\  B {\bf 576}, 173 (2003)
  [hep-ph/0307344].

\bibitem{reach3}
  M.~Ciuchini, E.~Franco, V.~Lubicz, F.~Mescia and C.~Tarantino,
  JHEP {\bf 0308}, 031 (2003)
  [hep-ph/0308029].



\bi{FG}  J.~S.~M.~Ginges and V.~V.~Flambaum,
  Phys.\ Rept.\  {\bf 397}, 63 (2004)
  [arXiv:physics/0309054].


\bi{PR}   M.~Pospelov and A.~Ritz,
  Annals Phys.\  {\bf 318}, 119 (2005)
  [arXiv:hep-ph/0504231].

\bi{Barr}  S.~M.~Barr,
  Phys.\ Rev.\  D {\bf 47}, 2025 (1993).
  
  \bi{BZ}  S.~M.~Barr and A.~Zee,
  Phys.\ Rev.\ Lett.\  {\bf 65}, 21 (1990)
  [Erratum-ibid.\  {\bf 65}, 2920 (1990)].

  
  \bi{Chang} D.~Chang, W.~Y.~Keung and T.~C.~Yuan,
  Phys.\ Lett.\  B {\bf 251}, 608 (1990).

  
  \bi{HPR} S.~J.~Huber, M.~Pospelov and A.~Ritz,
  Phys.\ Rev.\  D {\bf 75}, 036006 (2007)
  [arXiv:hep-ph/0610003].

  
  \bi{Pospelov} M.~Pospelov,
  Phys.\ Lett.\  B {\bf 530}, 123 (2002)
  [arXiv:hep-ph/0109044].

\bi{nedm} C.~A.~Baker {\it et al.},
  Phys.\ Rev.\ Lett.\  {\bf 97}, 131801 (2006)
  [arXiv:hep-ex/0602020].

\bibitem{MW}
  A.~V.~Manohar and M.~B.~Wise,
  Phys.\ Rev.\  D {\bf 74}, 035009 (2006)
  [arXiv:hep-ph/0606172].


  \bi{mike1} J.~M.~Arnold, B.~Fornal and M.~Trott,
  arXiv:1005.2185 [hep-ph].

\bibitem{mike2}
  J.~M.~Arnold, M.~Pospelov, M.~Trott and M.~B.~Wise,
  JHEP {\bf 1009}, 073 (2010)
  [arXiv:0911.2225 [hep-ph]].



\end{thebibliography}
\end{document}